\documentclass{ws-ijgmmp}
\usepackage[utf8]{inputenc}
\usepackage[T1]{fontenc}
\usepackage[colorlinks,hyperindex,plainpages=false]{hyperref}
\usepackage{cite}
\usepackage{amsmath}
\usepackage{amssymb}
\usepackage{amsfonts}
\usepackage{accents}
\allowdisplaybreaks

\begin{document}

\markboth{Ugur Camci}
{Noether symmetries of the minimal surface Lagrangian for G\"{o}del-type spacetimes}

%
\catchline{}{}{}{}{}
%

\title{Noether symmetries of the minimal surface Lagrangian for G\"{o}del-type spacetimes}

\author{Ugur Camci}

\address{Department of Chemistry and Physics, Roger Williams University, One Old Ferry Road, Bristol, RI 02809, USA\\
\email{ucamci@rwu.edu, ugurcamci@gmail.com}}

\maketitle

\begin{history}
\received{(Day Month Year)}
\revised{(Day Month Year)}
\end{history}

\begin{abstract}
We investigate the Noether symmetries of the minimal surface Lagrangian for four classes of metrics in G\"{o}del-type spacetimes. Then, calculating the Noether symmetries for all classes, namely, classes I, II, III and IV, we determine the conserved fields corresponding to each classes, allowing us to derive a comprehensive characterization of the minimal surface equations for G\"{o}del-type spacetimes.
\end{abstract}



\section{Introduction}
\label{INT}

G\"{o}del metric \cite{godel} which describes a homogeneous rotating universe containing closed time-like curves (CTCs), is one of the most interesting and well-known exact solution of Einstein's field equations, and is the best known example of causality violated universe model \cite{kramer}. Extending the G\"{o}del metric to G\"{o}del-type metrics involves generalizing the original solution to describe a broader class of spacetimes that retain some of the key features of the G\"{o}del universe, such as rotation and the presence of CTCs.
The G\"{o}del-type metrics should maintain rotational properties, meaning the metric should reflect a rotating universe. The rotational symmetry of G\"{o}del's metric comes from the existence of CTCs corresponding to circular orbits in specific coordinates. Furthermore these circular orbits have discussed by Raychaudhuri and Thakurta \cite{rayc}.
While exploring G\"{o}del-type metrics, one should consider their physical viability and whether they can represent realistic models or remain purely theoretical constructs.
The causality features of the G\"{o}del-type spacetimes are related with two independent parameters: $m$ and $\omega$. In papers \cite{rebo1, calvao1} it is shown that there exists only one non-causal region if $0 \leq m^2 < 4 \omega^2$, and there is no CTCs if $m^2 \geq 4 \omega^2$, in which the limiting case $m^2 = 4 \omega^2$ yields a completely causal and spacetime homogeneous G\"{o}del model, while there exists an infinite number of alternating causal and noncausal regions for $m^2 < 0$.

The G\"{o}del metrics are mainly interesting for their high degree of symmetry \cite{teix,rebo2,rebo3}. Since there exists a timelike Killing vector, both the original G\"{o}del metric and G\"{o}del-type metrics are stationary. However, a more detailed analysis of the Killing vectors shows, that those spacetimes are not static and also not isotropic, which is a feature due to the existence of a rotational axis giving rise to a rotational symmetry in the planes of constant $z$.
All classes of G\"{o}del-type spacetimes admit at least a $G_5$ group of motions. It has been shown that the group of motions is $G_7$, a maximal symmetry group of G\"{o}del-type spacetimes \cite{teix}, in a special case $m^2 = 4 \omega^2$.

The concept of minimal surfaces is closely related to the physics of surfaces and interfaces in various systems, and the mathematical framework for describing such surfaces involves the use of a minimal surface Lagrangian. One of the physical motivations behind the minimal surface Lagrangian is that
the physics of minimal surfaces is often associated with the minimization of surface energy. In systems with interfaces or surfaces, there is a tendency for the system to minimize its energy. This minimization is driven by the surface tension, which acts to reduce the area of the surface.
In mathematical terms, the minimal surface Lagrangian is often associated with the area functional, and the corresponding Euler-Lagrange (EL) equation describes surfaces with zero mean curvature. This mathematical formalism captures the physical behavior of surfaces and interfaces seeking to minimize their energy or surface area, providing a powerful tool for understanding various phenomena in physics.
To solve the EL equations, symmetries of the Lagrangian corresponding EL equations play an important role to find simple expressions for conservation laws. The Noether symmetries are associated with differential equations possessing a Lagrangian, and they are symmetries of the action integral. So, these symmetries concern all problems involving an action integral even if these problems do not involve the equations of motion. The geodesic equations of motion are the EL equations for arc length minimizing action. Therefore, they are important to study the evaluation of dynamical system \cite{qadir2012}. In the case of determination of the minimal surface area under constant volume in a given Riemannian space, the action integral involves the minimization of a surface and not of an arc length, as is the case with geodesic equations of motion.
Tsamparlis et al.\cite{tpq2015} have applied the Noether symmetry approach to a general Euclidian space, spaces of constant curvature and to Schwarzschild spacetime, and also showed how the Noether symmetries of minimal surface Lagrangian can be used to reduce the minimal surface equation form a partial differential equation (PDE) to an ordinary differential equation (ODE) in the Friedmann-Robertson-Walker (FRW) spacetime with dust source. In Ref.\cite{mhs2023} the authors found the Noether symmetries of minimal surfaces Lagrangian with fixed volume for plane-fronted gravitational wave (pp-wave) spacetimes. The Noether symmetry algebras admitted by wave equations on pp-waves with parallel rays are determined in Ref. \cite{js2016}. Noether symmetry analysis of some special classes of FRW universe and nonlinear wave equations in this geometry are performed by Camci \emph{et al.} \cite{camci2014}. It has been examined the the gravitational coupling of Klein-Gordon and Dirac fields to matter vorticity and spacetime torsion, in the background of G\"odel-type geometries \cite{kg-godel,marecki}, which admits the five Killing vectors.

The geodesic equations of motion for a space are important in the study of the evaluation of dynamical systems because of that the geometry of the background space determines the kinematics of the dynamical system in a unique manner. The symmetries of the corresponding EL equation is called Lie symmetries. Noether symmetries are also Lie symmetries, but the converse is not true in general. The relation of both the Lie and Noether symmetries of the geodesic equations for some spacetimes have been discussed by several authors \cite{tp2010,tp2011,cy2014}. The Noether symmetries are more important than Lie symmetries as it gives double reduction of the differential equations and provides conserved quantities.
The Noether symmetries can be used to reduce the order of the ODEs by considering the first integrals or the number of variable providing conserved quantities in the case of PDEs. In addition, it may be possible that one can linearize the nonlinear differential equations by means of the Noether equations. In recent years it has been provided a classification according to their geodesic Lagrangian considering Noether symmetry approach in the background of FRW \cite{tp2010}, Bertotti-Robinson like \cite{feroze1,feroze2,feroze3}, plane symmetric static \cite{feroze4} and pp-wave \cite{cy2014,cy2015}  spacetimes. The geodesic equations of motion for the general cylindrically symmetric stationary spacetimes together with their Dirac's constraint analysis and symplectic structure have been obtained, and integrated in Ref. \cite{ugur}.
The geodesic equations of motion in G\"odel-type spacetimes have been analyzed by several authors.
Firstly, Kunt \cite{kundt} was solved the geodesic equations for G\"odel's metric, where it was used the Killing vectors and corresponding constants of motion. Later,
Chandrasekhar and Wright \cite{chandra} presented an independent derivation of the solution for the geodesic equations of G\"odel's metric. A detailed study on geodesic motion in the original G\"{o}del's universe has been provided by Novello \emph{et al.} \cite{novello}.
The geodesic equations for the special case $m^2  = 4 \omega^2$ with seven isometries have been integrated by Rebou\c{c}as and Teixeira \cite{rebo4}. The geodesics of the Som-Raychaudhuri spacetime   \cite{som} have examined by Paiva \emph{et. al.} \cite{paiva}.
Grave \emph{et al.} \cite{grave} derived the analytical solution of the geodesic equations of G\"{o}del's universe for both particles and light, in which they have generalized the work of Kajari \emph{et. al.} \cite{kajari} on the solution of lightlike geodesic equations. Afterwards, Dautcourt \cite{daut} studied the lightcone of the G\"{o}del-type metrics by considering only the lightlike case. It is given a complete discussion for timelike and null geodesics of G\"{o}del-type spacetimes by Calv\~{a}o \emph{et al.} \cite{calvao2} and Gleiser \emph{et al.} \cite{gleiser}. Recently, Camci \cite{godel2014,camci2015} calculated the Noether symmetries for four classes of G\"{o}del-type metrics, and explicitly integrated the geodesic equations of motion by using the first integrals of corresponding classes.

This study is designed as follows. In the following section, we give a short review about G\"{o}del-type spacetimes and their properties. In section \ref{eqgm}, we present briefly the Noether symmetry approach for the minimal surface Lagrangian under constant volume and then apply this approach to the minimal surface Lagrangian of G\"{o}del-type spacetimes. In section \ref{soln}, we give solution of Noether symmetry equations and of the equations following from the conservation relation in each detail. Finally, our conclusion with a brief summary and discussions of finding is presented in Section \ref{conc}.

\section{G\"odel-type Spacetimes}
\label{fR}

The line element for the G\"{o}del-type spacetimes in cylindrical coordinates $x^a = \{ t,r,\phi,z \}$, $a = 0,1,2,3$, can be written as \cite{kramer,rayc}
\begin{equation}
ds^2 = \left[ dt + H(r) d\phi \right]^2 -dr^2 -D^2 (r) d\phi^2 -dz^2. \label{godel}
\end{equation}
It is found that the necessary and sufficient conditions for a G\"odel-type manifold to be spacetime
homogeneous (STH, hereafter) are \cite{rebo1,calvao1,teix,rebo2}
\begin{eqnarray}
& & \frac{D''}{D} = const \equiv m^2 , \label{cond1} \\& &
\frac{H'}{D} = const \equiv - 2\omega \label{cond2}
\end{eqnarray}
where prime denotes derivative with respect to the radial coordinate $r$. Throughout this paper we have used the following property
\begin{equation}\label{ozel1}
D^2 \left( \frac{D'}{D}  \right)' = -1,
\end{equation}
which is valid for STH G\"{o}del-type spacetimes only. The four-dimensional homogeneous Riemannian
G\"odel-type manifolds are locally characterized by two independent parameters $m^2$ and $\omega$: the pair of ($m^2,\omega$) identically specify locally equivalent manifolds.
All STH Riemannian manifolds endowed with a G\"{o}del-type spacetime \eqref{godel} are listed in Table \ref{Tab1}. We note that the scalar curvature $R$ of the G\"{o}del-type spacetimes becomes $R = 2 ( \omega^2 - m^2)$. Furthermore, if $m^2 = \omega = 0$, then the line element \eqref{godel} is
clearly Minkowskian. Therefore, this particular case has not been included in this study.

\begin{table}
\begin{tabular}{c|ccc}
\hline  \\
 & $H(r)$  & $D(r)$\quad &\quad $ m, \quad \omega$ \\
\hline \\
Class I & $\,\, \frac{2 \omega}{m^2} \left[ 1 - \cosh(mr) \right] \quad$ & $\frac{1}{m} \sinh(mr)\quad $ & $ m^2 > 0, \quad \omega \neq 0$ \\\\ \hline \\
Class II & $\,\, - \omega r^2 \quad$ & $ r $ & $m = 0, \quad \omega \neq 0$ \\\\ \hline \\
Class III & $ \frac{2 \omega}{\mu^2} \left[ \cos(\mu r) -1 \right]$ & $\frac{1}{\mu} \sin(\mu r) \quad $ & $ m^2 \equiv - \mu^2 < 0, \,\, \mu^2 >0, \, \omega \neq 0 $  \\\\ \hline \\
Class IV & $ 0$ & $\frac{1}{m} \sinh(mr) \quad $ & $m^2 > 0, \quad \omega = 0$ \\\\
 & $ 0$ & $ \frac{1}{\mu} \sin(\mu r) \quad $ & $m^2 \equiv - \mu^2 < 0, \quad \omega = 0$ \\
\hline
\end{tabular}
\caption{A list of all STH Rimannian manifolds endowed with a G\"{o}del-type spacetime \eqref{godel}, where the general solution of Eqs. \eqref{cond1} and \eqref{cond2} is written in each class. It is  referred to the manifolds of Class IV as degenerated G\"odel-type manifolds, since the cross term in the line element, related to the rotation $\omega$ in the G\"odel model, vanishes which means that we can make $H(r) = 0$ by a trivial coordinate transformation. }
\label{Tab1}
\end{table}

The group of conformal motions generated by a {\it conformal Killing vector} (CKV) field ${\bf K}$ is defined by $\pounds_{\bf K} g_{ab} = 2 \psi g_{ab}$, where  $\pounds_{\bf K}$ is the Lie derivative operator along the vector field ${\bf K}$, and $\psi = \psi (x^a)$ is a conformal factor.
The vector field ${\bf K}$ is an isometry or a Killing vector (KV) field if $\psi = 0$, and a homothetic vector (HV) if $\psi_{,a} = 0$. If $\psi_{;ab} \neq 0$, then the CKV field ${\bf K}$ is said to be {\it proper}, otherwise it is a special conformal Killing vector (SCKV) field when $\psi_{;ab} = 0$.
The set of all CKV (respectively SCKV, HKV and KV) form a finite-dimensional Lie algebra denoted by $\mathcal{C}$ (respectively $\mathcal{S}, \mathcal{H}$ and $\mathcal{G}$).

It is proved by Rebou\c{c}as \emph{et al.} \cite{rebo3} that the four-dimensional homogeneous Riemannian G\"odel-type manifolds admit a group of isometry $G_r$ with

{\bf (i)} $r = 5$ in classes I (where $m^2 \neq 4 w^2$), II and III;

{\bf (ii)} $r = 6$ in class IV;

{\bf (iii)} $r = 7$ in the special case of class I, where $m^2 = 4 \omega^2$.
\\
The KV fields of the classes I-IV of STH G\"{o}del-type spacetimes \eqref{godel} have been determined as follows.
In case (i), the {\it five KVs} ${\bf K}_1,..., {\bf K}_5$ for class I are obtained as
\begin{eqnarray}
& & {\bf K}_1 = \partial_t, \quad {\bf K}_2 = \partial_z, \quad
{\bf K}_3 = \frac{2 \omega}{m} \partial_t - m \partial_{\phi},
\nonumber \\ & & {\bf K}_4 = - \frac{H}{D} \sin \phi \partial_t + \cos \phi
\partial_r - \frac{D'}{D} \sin \phi \partial_{\phi}, \label{KV-case-i} \\
& & {\bf K}_5 = - \frac{H}{D} \cos \phi \partial_t - \sin \phi
\partial_r - \frac{D'}{D} \cos \phi \partial_{\phi}, \nonumber
\end{eqnarray}
For class III in case (i), where $m^2 \equiv - \mu^2 <0, \mu^2 >0$ and $\omega \neq 0$, it follows that the KVs ${\bf K}_1, {\bf K}_2, {\bf K}_4, {\bf K}_5$ are the same form as given the above, but only ${\bf K}_3$ has the form $(2 \omega / \mu )\partial_t + \mu \partial_{\phi}$. For class II of case (i), where $H(r) = - \omega r^2$ and $D(r) = r$, the {\it five KVs} ${\bf K}_1,..., {\bf K}_5$ are given by
\begin{eqnarray}
& & {\bf K}_1 = \partial_t, \quad {\bf K}_2 = \partial_z, \quad
{\bf K}_3 = \partial_{\phi}, \,\,  {\bf K}_4 = - \omega \, r \, \sin\phi \partial_t - \cos\phi \partial_r + \frac{1}{r} \sin\phi \partial_{\phi}, \nonumber \\ & &
{\bf K}_5 = - \omega \, r \, \cos\phi \partial_t + \sin\phi
\partial_r + \frac{1}{r} \cos\phi \partial_{\phi} \, . \label{KV-class-II}
\end{eqnarray}
For class IV in case (ii), the {\it six KVs} ${\bf K}_1,..., {\bf K}_6$ are
\begin{eqnarray}
& & {\bf K}_1 = \partial_t, \quad {\bf K}_2 = \partial_z, \quad
{\bf K}_3 = z \, \partial_t + t \partial_z, \nonumber
\quad {\bf K}_4 = \cos \phi \partial_r - \frac{D'}{D} \sin \phi \partial_{\phi},  \\
& & {\bf K}_5 = - \sin \phi \partial_r - \frac{D'}{D} \cos \phi \partial_{\phi}, \quad {\bf K}_6 = \partial_{\phi} \, .  \label{KV-case-ii}
\end{eqnarray}
Finally, for the special class I in case (iii), where $m^2 = 4 \, \omega^2$, i.e. $m = + 2 \,\omega$, the {\it seven KVs} ${\bf K}_1,...,{\bf K}_7$ are given by
\begin{eqnarray}
& & {\bf K}_1 = \partial_t, \quad {\bf K}_2 = \partial_z, \quad
{\bf K}_3 = \partial_t - m \partial_{\phi}, \nonumber
\\ & & {\bf K}_4 = - \frac{H}{D} \sin \phi \partial_t + \cos \phi \partial_r - \frac{D'}{D} \sin\phi \partial_{\phi},  \nonumber \\
& & {\bf K}_5 = - \frac{H}{D} \cos \phi \partial_t - \sin \phi \partial_r - \frac{D'}{D} \cos \phi \partial_{\phi}, \label{KV-case-iii}  \\&
& {\bf K}_6 = -\frac{H}{D} \cos (m t + \phi) \partial_t + \sin (m t + \phi) \partial_r + \frac{1}{D} \cos (m t + \phi) \partial_{\phi},
\nonumber \\& & {\bf K}_7 = -\frac{H}{D} \sin (m t + \phi) \partial_t - \cos (m t + \phi) \partial_r + \frac{1}{D} \sin (m t + \phi) \partial_{\phi} \, . \nonumber
\end{eqnarray}

It is known that both the original G\"odel metric and the STH  G\"odel-type spacetimes does not admit HVs \cite{hall-costa,melfo}. The proper CKVs and complete conformal algebra of a G\"odel-type spacetime have been computed in Ref. \cite{tsamparlis}. The Ricci collineations and the matter collineations of STH G\"odel-type spacetimes have been obtained in Refs. \cite{melfo} and \cite{cmc-sharif}, respectively. In this work, we aim to give a complete classification for STH G\"{o}del-type spacetimes according to the Noether symmetries of their minimal surface Lagrangian.

\section{Noether Symmetries of the Minimal Surface Lagrangian} \label{eqgm}

First, we will explain Noether symmetry approach for a first-order Lagrangian $L (x^i, q^{\alpha}, q^{\alpha}_i )$ as the following. Let $x^i$ and $q^{\alpha}$ be $n$-independent and $m$-dependent variables, respectively. The total derivative operator is given by
\begin{equation}
  D_i = \partial_{x^i} + q^{\alpha}_i \partial_{q^{\alpha}} + q^{\alpha}_{i j} \partial_{q^{\alpha}_j} + ...,
\end{equation}
where $q^{\alpha}_i = D_i (q^{\alpha}), q^{\alpha}_{i j} = D_i D_j ( q^{\alpha}),...$ represent the derivatives of $q^{\alpha}$ with respect to $x^i$. Then the Euler-Lagrange equations corresponding to the Lagrangian $L (x^i, q^{\alpha}, q^{\alpha}_i )$ are
\begin{equation}
  \frac{\partial L}{\partial q^{\alpha} } - \frac{\partial}{\partial x^i}\left( \frac{\partial L}{\partial q^{\alpha}_i } \right) = 0, \label{EL-eq}
\end{equation}
where $i,j,...=1,...,n$ and $\alpha, \beta, ... = 1,...,m$. The infinitesimal symmetry generator
\begin{equation}
  {\bf X} = \xi^i (x^k, q^{\beta} ) \partial_{x^i} + \eta^{\alpha} (x^k, q^{\beta} ) \partial_{q^{\alpha}}  \label{NS-gen}
\end{equation}
is called a Noether symmetry (NS) of the Lagrangian $L$ if there exits a vector-valued gauge function ${\bf A} = A^i (x^k, q^{\beta} ) \partial_{x^i}$ such that the following condition is satisfied,
\begin{eqnarray}
  & & {\bf X}^{[1]} L + L D_i ( \xi^i ) = D_i ( A^i ),  \label{ns-cond}
\end{eqnarray}
where ${\bf X}^{[1]}$ is the first-order prolongation vector field defined by ${\bf X}^{[1]} = {\bf X} + \eta^{\alpha}_i \partial_{ q^{\alpha}_i } $ with $ \eta^{\alpha}_i = D_i \eta^{\alpha} - q^{\alpha}_j D_i ( \xi^j )$. The components of the conserved vector field ${\bf T} = T^i \partial_{x^i}$ which is also called Noether flow satisfying $D_ i T^i = 0$, is given by \cite{noether}
\begin{eqnarray}
  & & T^i = \xi^i L + \left( \eta^{\alpha} - \xi^j q^{\alpha}_j  \right) \frac{\partial L}{\partial q^{\alpha}_i } - A^i \, . \label{con-law}
\end{eqnarray}

Now, we will determine the Noether symmetries of the minimal surface Lagrangian $L$ for the G\"{o}del-type spacetime by applying Noether symmetry approach summarized above. Thus, after a simple rearrangement, one can also write the G\"{o}del-type spacetime \eqref{godel} as
\begin{equation}
  ds^2 = h_{i j} dx^i dx^j - dz^2 \, ,  \label{godel2}
\end{equation}
where $i,j,...= 1,2,3$, and the metric $h_{i j}$ is of the form
\begin{equation}
h_{i j} = \left(
  \begin{array}{ccc}
    1 & \, \ 0 & H \\
    0 & -1 & 0 \\
    H & \,\ 0 & \, \ H^2 - D^2 \\
  \end{array}
\right)
\end{equation}
which yields $h= \det ( h_{ij} ) = D^2$.  Therefore, the number of independent and dependent variables are $n=3$ and $m = 1$, respectively, that means $x^i = \{ t, r, \phi\}$ and $q = \{ z = f (t,r,\phi) \}$. We note that the metric \eqref{godel2} is $(1 + 3)$-decomposable because it is {\it not} a function of $z=f(t,r,\phi)$, which gives rise to the metric \eqref{godel2} that admits the gradient KV $\partial_z$.
The Lagrangian of minimal surface for a 3-area enclosing a constant volume is given by \cite{tpq2015}
\begin{eqnarray}
  & & L = \sqrt{ |h| + |h| h^{i j} f_{,i} f_{,j}  } + \lambda \int{\sqrt{|h|}} d f \, , \label{lagr-1}
\end{eqnarray}
which gives
\begin{eqnarray}
  & & L = \sqrt{ D^2 + (D^2 - H^2) f_{,t}^2 + 2 H f_{,t} f_{,\phi} - D^2 f_{,r}^2 - f_{,\phi}^2 } + \lambda \, D f(t,r,\phi) \, , \label{lagr-godel2}
\end{eqnarray}
for the G\"{o}del-type spacetime \eqref{godel2}.
The minimal surface equation for this spacetime which results from the Lagrangian \eqref{lagr-godel2} is a second-order partial differential equation in which independent variables are $x^i = \{ t,r,\phi \}$. Then, one may obtain the Euler-Lagrange equation of minimal surface for the Lagrangian \eqref{lagr-godel2} by using \eqref{EL-eq} as
\begin{eqnarray}
    & & (D^2 - H^2) \left( 1 - f^2_{,r} \right) f_{,tt} + 2 \left[ \left( 1 - f^2_{,r} \right) H  + f_{,t} f_{,\phi} \right] f_{,t \phi}  \nonumber \\ & & \,\, - \left( 1 + f^2_{,t} - f^2_{,r} \right) f_{,\phi \phi}  - \left[ D^2 \left( 1 + f_{,t}^2 \right) + 2 H f_{,t} f_{,\phi} - H^2 f^2_{,t} - f^2_{,\phi} \right] f_{,rr}  \nonumber \\ & & \,\,  + 2 f_{,r} \left[ (D ^2 - H^2) f_{,t} + H f_{,\phi}  \right] f_{,t r}   + 2 f_{,r} ( H f_{,t} - f_{,\phi} ) f_{,r \phi} - \frac{\lambda}{D} F^3  \nonumber \\ & &  \,\, + f_{,r} \left[ \left( D D' ( 3 - \frac{2 H^2}{D^2} ) - H H' \right) f^2_{,t} + H \left( \frac{H'}{H} + \frac{4 D'}{D} \right) f_{,t} f_{,\phi} - D D' \left( 1 +  3 f^2_{,r} \right) \right]  = 0, \quad \qquad \label{EL-eq-lagr}
\end{eqnarray}
where $F$ is defined by
\begin{equation}
 F \equiv \sqrt{ D^2 + (D^2 - H^2) f_{,t}^2 + 2 H f_{,t} f_{,\phi} - D^2 f_{,r}^2 - f_{,\phi}^2 } \, . \label{defn-F}
\end{equation}
Using this definition, the Lagrangian \eqref{lagr-godel2} becomes $L = F + \lambda \, D f(t,r,\phi)$. Introducing the momenta $P^i = \frac{\partial L}{\partial f_{,i}}$, we have
\begin{equation}
  P^t = \frac{1}{F} \left[ (D^2 - H^2) f_{,t} + H f_{,\phi} \right] \, , \quad P^r =- \frac{D^2}{F} f_{,r} \, \, , \quad P^{\phi} = \frac{1}{F} \left( H f_{,t} - f_{,\phi} \right) \, . \label{momenta}
\end{equation}
One can define the Hamiltonian function corresponding to the minimal surface Lagrangian such as $\mathcal{H}_L = f_{,i} P^i - L$, which becomes
\begin{equation}
  \mathcal{H}_L = - D \left( \frac{D}{F} + \lambda \, f \right) \, , \label{hamilton}
\end{equation}
by using the components of the momenta $P^i$ in \eqref{momenta}.

Now, let us consider the Noether symmetry generator
\begin{equation}\label{vecf}
{\bf X} = \eta \, \partial_f + \xi^1 \partial_t + \xi^2 \partial_r  + \xi^3 \partial_{\phi}
\end{equation}
where $\eta, \xi^1, \xi^2$ and $\xi^3$ are depend on $x^i = \{ t,r,\phi \}$ and $z= f (t, r, \phi)$. The first extension of the above vector field is given by
\begin{equation}\label{prol}
{\bf X^{[1]}} = {\bf X}+ \eta'_t \, \partial_{f_{,t}} + \eta'_r \, \partial_{f_{,r}}
+ \eta'_{\phi} \, \partial_{f_{,\phi} } \, ,
\end{equation}
in which $\eta'_t  = D_{t} \eta - f_{,j} D_{t} \xi^j$, $\eta'_r  = D_{r} \eta - f_{,j} D_{r} \xi^j$ and $\eta'_{\phi}  = D_{\phi} \eta - f_{,j} D_{\phi} \xi^j$,
where $D_i$'s are the operator of total differentiation with respect to $x^i$ such that $D_t = \partial_t + f_{,t} \partial_f$, $D_r = \partial_r + f_{,r} \partial_f$ and $D_{\phi} = \partial_{\phi} + f_{,\phi} \partial_f$. Then, it follows from Eq. \eqref{ns-cond} that for the minimal surface Lagrangian \eqref{lagr-godel2}, the Noether symmetry equations yield {\emph{14} partial differential equations:
\begin{eqnarray}
  & & \lambda D ( \xi^1 + f \xi^1_{,f} ) - A^1_{,f} = 0, \quad \lambda D ( \xi^2 + f \xi^2_{,f} ) - A^2_{,f} = 0, \,\, \lambda D ( \xi^3 + f \xi^3_{,f} ) - A^3_{,f} = 0, \nonumber \\ & & (D^2 - H^2)\eta_{,t} + H \eta_{,\phi} + D^2 \xi^1_{,f} = 0, \quad \eta_{,r} - \xi^2_{,f} = 0, \quad H \eta_{,t} - \eta_{,\phi} + D^2 \xi^3_{,f} = 0, \nonumber \\ & & \frac{D'}{D} \xi^2 + \eta_{,f} + \xi^1_{,t} +  \xi^3_{,\phi} = 0, \quad  \eta_{,f} - \xi^2_{,r} = 0 \, , \quad \lambda D \eta  - A^1_{,t} - A^2_{,r} - A^3_{,\phi} = 0 \, , \nonumber \\ & & \frac{D'}{D} \xi^2 - \xi^2_{,r} - H \xi^3_{,t} + \xi^3_{,\phi} = 0, \,\,  D^2 \xi^1_{,r} - (D^2 - H^2) \xi^2_{,t} - H \xi^2_{,\phi} = 0 \, , \,\,  \label{ns-eqs}  \\ & & \left( \frac{H'}{H} - \frac{2 D'}{D} \right) \xi^2 + 2 \xi^2_{,r} - \xi^1_{,t} + \frac{1}{H} \xi^1_{,\phi} - \xi^3_{,\phi} - \frac{(D^2-H^2)}{H} \xi^3_{,t} = 0 , \nonumber \\ & & H^2 \left( \frac{D'}{D} - \frac{H'}{H} \right) \xi^2 + (D^2 - H^2) ( \xi^2_{,r} - \xi^1_{,t} ) - H \xi^1_{,\phi} = 0 \, , \xi^2_{,\phi} - H \xi^2_{,t} + D^2 \xi^3_{,r} = 0, \nonumber
\end{eqnarray}
where the subscripts with comma denotes partial derivatives. It is noted here that the set of all NSs form a finite dimensional Lie algebra denoted by $\mathcal{N}$.
The significance of NS is clearly comes from the fact that if ${\bf X}$ is the Noether symmetry corresponding to the Lagrangian $L(t,r,\phi,f,f_{,t},f_{,r},f_{,\phi})$ given by \eqref{lagr-godel2}, then
\begin{equation}
T^i = \xi^i L + \left(\eta -\xi^j f_{,j} \right) \frac{\partial
L}{\partial f_{,i}} - A^i \, , \label{con-law-2}
\end{equation}
is a \emph{vector-valued quantity} associated with the vector field ${\bf X}$. Then the Noether current ${\bf T} = T^1 \partial_t + T^2 \partial_r + T^3 \partial_{\phi}$, where the components satisfies the relation
\begin{equation}
   D_i T^i = 0 \quad \Longleftrightarrow  \quad D_t T^1 + D_r T^2 + D_{\phi} T^3 = 0 \, , \label{con-law-3}
\end{equation}
is conserved along with the solutions of minimal surface equation \eqref{EL-eq-lagr}. Corresponding to each of ${\bf X}$, there exists a conserved vector ${\bf T}$ and then one can write the components of the Noether conserved vector from \eqref{con-law-2} as
\begin{eqnarray}
  & & T^1 = \xi^1 L + \left( \eta - f_{,t} \xi^1 - f_{,r} \xi^2 - f_{,\phi} \xi^3 \right) \frac{\partial L}{\partial f_{,t}} - A^1 \, , \nonumber \\ & &  T^2 = \xi^2 L + \left( \eta - f_{,t} \xi^1 - f_{,r} \xi^2 - f_{,\phi} \xi^3 \right) \frac{\partial L}{\partial f_{,r}} - A^2 \, , \label{con-law-comp}  \\ & & T^3 = \xi^3 L + \left( \eta - f_{,t} \xi^1 - f_{,r} \xi^2 - f_{,\phi} \xi^3 \right) \frac{\partial L}{\partial f_{,\phi}} - A^3 \, . \nonumber
\end{eqnarray}

Recently the Noether symmetries of minimal surface Lagrangian for some spacetimes have been calculated, and classified according to their symmetry generators \cite{qadir2012,tpq2015,mhs2023}. In this study, we investigate the NSs of minimal surface Lagrangian for the G\"{o}del-type spacetimes. The general solution to the above NS equations is introduced in the next section for each classes I-IV of the G\"{o}del-type spacetimes.

\section{The Solution of Noether Symmetry Equations } \label{soln}

After some algebra, we have calculated the general solution to Eqs. \eqref{ns-eqs} in order to get NSs of minimal surface Lagrangian for each of the classes  I, II, III and IV as the following.

\subsection{Classes I, II and III}

The components of NS vector field ${\bf X}$ and the vector-valued gauge function ${\bf A}$ for classes I and III with the condition $m^2 \neq \omega^2$ are found as
\begin{eqnarray}
& & \eta = c_1, \quad \xi^1 = \frac{H}{D} (c_2 \cos \phi - c_3 \sin\phi) + c_4 , \,\, \nonumber \\& & \xi^2 =  c_2 \sin\phi + c_3 \cos\phi, \quad \xi^3 = \frac{D'}{D} (c_2 \cos\phi - c_3\sin\phi) + c_5 , \nonumber \\ & & A^1 = \lambda \, f \left[ H \, ( c_2 \cos\phi - c_3 \sin\phi ) + c_4 D  \right] + \Phi^1(t,r,\phi), \label{ns-classes123} \\ & &  A^2 = \lambda \, f \, D \, (c_2 \sin\phi + c_3 \cos\phi) + \Phi^2 (t,r,\phi), \nonumber \\ & & A^3 = c_1 \lambda \phi D + \lambda \, f \left[ D' ( c_2 \cos\phi - c_3 \sin\phi ) + c_5 D \right] - \int{(\Phi^1_{,t} + \Phi^2_{,r} ) d \phi } + \Phi^3 (t,r), \nonumber
\end{eqnarray}
where $c_1, ..., c_5$ are constant parameters, $\Phi^i$'s are integration functions.
Hence, we have {\it five NSs} for classes I and III, which can be stated by means of the KVs given in \eqref{KV-case-i} for class I such as
\begin{equation}
  {\bf X}_1 = {\bf K}_1 \, , \,\, {\bf X}_2 = {\bf K}_2  \, , \,\, {\bf X}_3 = \partial_{\phi} =  \frac{2 \omega}{m^2} {\bf K}_1 - \frac{1}{m} {\bf K}_3 \, , \,\, {\bf X}_4 = {\bf K}_4 \, , \,\, {\bf X}_5 = {\bf K}_5 \, , \label{ns-class-I}
\end{equation}
together with the corresponding gauge vectors
\begin{eqnarray}
  & & {\bf A}_1 = \lambda \, f D \, {\bf X}_1 ,  {\bf A}_2 = \lambda \phi D \, {\bf X}_3 ,  {\bf A}_3 = \lambda \, f D \, {\bf X}_3 ,  {\bf A}_4 = \lambda \, f D \, {\bf X}_4 , {\bf A}_5 = \lambda \, f  D {\bf X}_5 \, , \qquad \label{gauge-i-1-5}
\end{eqnarray}
where ${\bf K}_1, ...,{\bf K}_5$ are the KVs for classes I and III given in \eqref{KV-case-i}. Also, it is seen from \eqref{ns-classes123} that for any NS of these classes there exists a generic gauge vector functions such as
\begin{equation}
  {\bf A}_0 = \Phi^1 (t,r,\phi) \partial_t + \Phi^2 (t,r,\phi) \partial_r + \left[ \Phi^3 (t,r) - \int{ (\Phi^1_{,t} + \Phi^2_{,r} ) d \phi } \right] \partial_{\phi} \, .
\end{equation}
In class III, the NSs ${\bf X}_1, {\bf X}_2, {\bf X}_4$ and ${\bf X}_5$ are the same form as for class I described in \eqref{ns-class-I}, but the NS ${\bf X}_3$ becomes ${\bf X}_3 = \partial_{\phi} = -(2 \omega/ \mu^2) {\bf K}_1 + (1/\mu) {\bf K}_3$.
The Lie algebra of the NSs \eqref{ns-class-I} has the following non-vanishing commutators:
\begin{eqnarray}
& & \left[ {\bf X}_3, {\bf X}_4 \right] = {\bf X}_5, \quad
\left[ {\bf X}_3, {\bf X}_5 \right] = - {\bf X}_4, \quad \left[
{\bf X}_4, {\bf X}_5 \right] = 2 \omega {\bf X}_1 - m^2 {\bf X}_3 . \nonumber
\end{eqnarray}
The conserved vector fields following from Eq.\eqref{con-law-2} for the Noether symmetries ${\bf X}_1,...,{\bf X}_5$ of classes I and III read, respectively,
\begin{eqnarray}
  & & {\bf T}_1 = F \, {\bf X}_1 - \frac{f_{,t}}{F} {\bf T}_0 \, , \quad {\bf T}_2 = - \lambda \phi D \, {\bf X}_3 + \frac{1}{F} {\bf T}_0 \, ,  \quad {\bf T}_3 = F \, {\bf X}_3 - \frac{f_{,\phi}}{F} {\bf T}_0\, , \label{cnserv-i-1} \\ & & {\bf T}_4 = F \, {\bf X}_4 + \frac{1}{F} \left[ \frac{H}{D} f_{,t} \sin\phi - f_{,r} \cos\phi + \frac{D'}{D} f_{,\phi} \sin\phi \right] {\bf T}_0 \, , \label{cnserv-i-2} \\ & & {\bf T}_5 = F \, {\bf X}_5 + \frac{1}{F} \left[ \frac{H}{D} f_{,t} \cos\phi + f_{,r} \sin\phi + \frac{D'}{D} f_{,\phi} \cos\phi \right] {\bf T}_0 \, , \label{cnserv-i-3}
\end{eqnarray}
where the function $F$ has the form as given in \eqref{defn-F} and we have defined ${\bf T}_0$ as follows:
\begin{equation}
  {\bf T}_0 =  [(D^2 - H^2) f_{,t} + H f_{,\phi}]\partial_t  - D^2 f_{,r} \partial_r + ( H f_{,t} - f_{,\phi}) \partial_{\phi} \, . \label{T0}
\end{equation}
Ultimately, the conservation relation $D_i T^i = 0$ for the conserved vector fields \eqref{cnserv-i-1}-\eqref{cnserv-i-3} yields
\begin{eqnarray}
   & & f_{,r} \, f_{,t} \left( \frac{D^2}{F} \right)' = 0 \, , \quad  f_{,r} f_{,\phi} \left( \frac{D^2}{F} \right)' = 0 \, , \quad f_{,r} \left( \frac{D^2}{F} \right)' + \lambda \, D = 0 \, , \label{cv-c13-1} \\ & &   f_{,r} f_{,t} \left[ \left( \frac{H D}{F} \right)' - \frac{H}{F}  \right] + f_{,r} f_{,\phi} \left[ \left( \frac{D D'}{F} \right)' + \frac{1}{F} \right] = 0 \, , \label{cv-c13-2} \\ & & F' - \frac{D'}{D} F + f_{,t}^2 \frac{H^2}{F D} + f_{,r}^2 \left( \frac{D^2}{F} \right)' - f_{,\phi}^2 \frac{D'}{F D}  + f_{,t} f_{,\phi} \frac{H (D' -1)}{F D} = 0 \, . \label{cv-c13-3}
\end{eqnarray}
Here, we have three different cases for classes I and III following from the above equations:
\begin{eqnarray}
  & & (a) \, \left( \frac{D^2}{F} \right)' = 0 \, , \, \lambda = 0; \quad (b)\, f_{,r} = 0 \, , \, \lambda = 0;\quad (c)\, f_{,t} = f_{,\phi} = 0\, , \, \lambda \neq 0 \, . \label{cases-abc}
\end{eqnarray}
In case ({\it a}), the equations in \eqref{cv-c13-1} are identically satisfied, and the condition $( D^2 / F)' = 0$ has a solution as $F = D^2 / h (f_{_{,t}}\, , f_{_{,r}}\, , f_{_{,\phi}})$, where $h (f_{_{,t}}\, , f_{_{,r}}\, , f_{_{,\phi}})$ is an integration function. Then, Eq.\eqref{cv-c13-2} becomes $f_{,r} f_{,t} \left[ (H/D)' - H/D^2 \right] = 0$ if $f_{,t}, f_{,r} \neq 0$, which also identically satisfied for classes I and III. For this case, the Eq. \eqref{cv-c13-3} yields
\begin{equation}
  D^4 D' + h^2 \left[ H^2 f_{,t}^2 - D' f_{,\phi}^2 + H (D'-1) f_{,t} f_{,\phi} \right] = 0 \, .
\end{equation}
In case ({\it b}), Eqs. \eqref{cv-c13-1} and \eqref{cv-c13-2} are directly identities. Further, we can write Eq.\eqref{cv-c13-3} as
\begin{equation}
  (F^2)' - \frac{2 D'}{D} F^2 + \frac{2 H f_{,t}}{D} \left[ H f_{,t}  + (D' -1) f_{,\phi} \right] - \frac{2 D'}{D} f_{,\phi}^2  = 0 \, , \label{cv-c13-3-2}
\end{equation}
which gives $f_{,\phi} = - H f_{,t}$ using $F = \sqrt{ D^2 + (D^2 - H^2) f_{,t}^2 + 2 H f_{,t} f_{,\phi} - f_{,\phi}^2}$. Since $f_{,r} = 0$ (i.e., $f= f(t,\phi)$) in this case, the relation $f_{,\phi} = - H f_{,t}$ yields $H' = 0$, i.e.  $\omega = 0$. So the case ({\it b}) reduces to the class IV. Therefore this case should be excluded as a possibility for the classes I and III.
In case ({\it c}), $f= f(r)$ and one of the relations in \eqref{cv-c13-1} remains such that \begin{equation}
   f_{,r} \left( \frac{D^2}{F} \right)' + \lambda \, D = 0 \, . \label{cv-c13-1-2}
\end{equation}
Additionally, Eq. \eqref{cv-c13-2} is an equality, and Eq.\eqref{cv-c13-3} becomes
\begin{equation}
  F' - \frac{D'}{D} F - \lambda D f_{,r} = 0 \, , \label{cv-c13-3-3}
\end{equation}
regarding \eqref{cv-c13-1-2}. The equation \eqref{cv-c13-3-3} is a linear first-order ordinary differential equation in terms of $F$, and has a solution as $F= D ( \lambda f(r) + b_1 )$, where $b_1$ is a constant of integration. Using this solution of $F$, the Eq.\eqref{cv-c13-1-2} yields
\begin{equation}
  f_{,r}^2 - ( \lambda f + b_1) f_{,r} \frac{D'}{\lambda D} - \left( \lambda f + b_1 \right)^2 = 0 \, . \label{cv-c13-1-3}
\end{equation}
Further, one can easily see from the definition of $F$ that for this case $F = D \sqrt{1 - f_{,r}^2}\,$, which gives directly that
\begin{equation}
  \sqrt{ 1 - f_{,r}^2} - ( \lambda f + b_1) = 0 \, . \label{cv-c13-1-4}
\end{equation}
Thus, introducing a new variable $U(r) = \lambda f + b_1$ and taking into account the relation \eqref{cv-c13-1-4} in \eqref{cv-c13-1-3}, it follows the following differential equation
\begin{equation}
  \frac{D'}{D} \left( U^2 \right)' + 2 \lambda^2 (1 - 2 U^2 ) = 0 \, , \label{cv-c13-1-5}
\end{equation}
which has a general solution as
\begin{equation}
  U(r)^2 = \frac{1}{2} + b_2 \left( D' \right)^{-\frac{4 \lambda^2}{m^2}} \, , \label{sol-U}
\end{equation}
where $b_2$ is an integration constant. Then, after substitution \eqref{sol-U} into $U(r) = \lambda f$ in which we  take $b_1 = 0$ without loss of generality, one can get for $f(r)$ that
\begin{equation}
  f(r) = \frac{1}{\lambda} \sqrt{ \frac{1}{2} + b_2 \left( D' \right)^{-\frac{4 \lambda^2}{m^2}} } \, . \label{sol-f(r)}
\end{equation}
Also, we find from Eq.\eqref{cv-c13-1-4} that $\lambda^2 = - m^2/4$ and $b_2 = \pm 1/2$ for class I, and $\lambda^2 = \mu^2/4$ and $b_2 = \pm 1/2$ for class III. Therefore, the final form of $f(r)$ in case ({\it c}) becomes
\begin{equation}
  f(r) =
\left\{
  \begin{array}{ll}
    \frac{\sqrt{2}}{m} \sqrt{-1 \pm \cosh (m r) }  & \quad \hbox{for class I;} \\ \\
    \frac{\sqrt{2}}{\mu} \sqrt{1 \pm \cos(\mu r)}  & \quad \hbox{for class III.}
  \end{array}
\right. \label{sol-fr-2}
\end{equation}

In the special class I case, where $m^2 = 4 \omega^2$, i.e. $ \omega = + m/2$, we found that the components $\eta, \xi^1, \xi^2, \xi^3$ of the Noether symmetry generator ${\bf X}$, and the gauge vector field components $A^1,A^2$ and $A^3$ are
\begin{eqnarray}
& & \eta = c_1, \quad \xi^1 = -\frac{H}{D} \left[ c_2 \sin\phi - c_3 \cos\phi + c_4 \sin ( m t + \phi) + c_5 \cos( m t +\phi )  \right] + c_6 \, \nonumber  \\& & \xi^2 = c_2 \cos\phi + c_3 \sin\phi  - c_4 \cos (m t + \phi) + c_5 \sin (m t + \phi)\,  , \nonumber \\ & &  \xi^3 = -\frac{D'}{D} ( c_2 \sin\phi - c_3\cos\phi ) + \frac{1}{D} \left[ c_4 \sin (m t + \phi) + c_5 \cos (m t + \phi) \right]  +  c_7 \, ,  \nonumber \\ & & A^1 = - \lambda  f H \left[ -c_2 \sin\phi + c_3 \cos\phi  + c_4 \sin (m t + \phi) + c_5 \cos ( m t + \phi ) \right] \nonumber \\ & & \qquad \qquad + c_6 \lambda f D + \Phi^1(t,r,\phi) \, ,  \\   & &  A^2 = \lambda \, f D \, \left[ c_2 \cos\phi + c_3 \sin\phi - c_4 \cos ( m t + \phi) + c_5 \sin ( m t + \phi )  \right] + \Phi^2 (t,r,\phi),  \nonumber \\ & & A^3 = \lambda \, f  \left[ D' ( -c_2 \sin\phi + c_3 \cos \phi ) + c_4 \sin ( m t + \phi ) + c_5 \cos ( m t + \phi ) + c_7 D \right] \nonumber \\ & &  \qquad \quad + c_1 \lambda \phi D - \int{(\Phi^1_{,t} + \Phi^2_{,r} ) d \phi } + \Phi^3 (t,r) \, . \nonumber
\end{eqnarray}
Then, one finds that there are {\it seven} NSs of minimal surface Lagrangian which are given by
\begin{eqnarray}
  & & {\bf X}_1 = {\bf K}_1 \, , \quad {\bf X}_2 = {\bf K}_2  \, , \quad {\bf X}_3 = \partial_{\phi} =  \frac{1}{m} \left( {\bf K}_1 - {\bf K}_3 \right) \, ,  \label{ns-class-I-s1} \\ & & {\bf X}_4 = {\bf K}_4 \, , \quad {\bf X}_5 = {\bf K}_5 \, , \quad {\bf X}_6 = {\bf K}_6 \, , \quad {\bf X}_7 = {\bf K}_7 \, . \label{ns-class-I-s2}
\end{eqnarray}
The gauge vectors of ${\bf X}_1, {\bf X}_2$ and ${\bf X}_3$ are the same given in \eqref{gauge-i-1-5}, and the gauge vectors for the NSs ${\bf X}_4, {\bf X}_5, {\bf X}_6$ and ${\bf X}_7$ yield
\begin{eqnarray}
    & & {\bf A}_4 = \lambda \, f D \, {\bf K}_4 \, , \quad {\bf A}_5 =  \lambda \, f D \, {\bf K}_5  \, , \quad {\bf A}_6 = \lambda \, f D \, {\bf K}_6 \, , \quad {\bf A}_7 = \lambda \, f D \, {\bf K}_7\, . \label{gauge-i-s}
\end{eqnarray}
The corresponding Lie algebra has the following non-vanishing commutators:
\begin{eqnarray}
& & \left[ {\bf X}_1, {\bf X}_6 \right] = - m {\bf X}_7, \,\,
\left[ {\bf X}_1, {\bf X}_7 \right] = m {\bf X}_6, \quad \left[
{\bf X}_3, {\bf X}_4 \right] =  {\bf X}_5 , \quad \left[ {\bf X}_3, {\bf X}_5 \right] = - {\bf X}_4,  \nonumber \\& & \left[ {\bf X}_3, {\bf X}_6 \right] = - {\bf X}_7, \,\, \left[ {\bf
X}_3, {\bf X}_7 \right] =  {\bf X}_6, \, \left[ {\bf
X}_4, {\bf X}_5 \right] = m {\bf X}_1 - m^2 {\bf X}_3, \, \left[ {\bf
X}_6, {\bf X}_7 \right] = m {\bf X}_1 . \nonumber
\end{eqnarray}
The conserved vector fields of this special class I for ${\bf X}_1,{\bf X}_2, {\bf X}_3, {\bf X}_4$ and ${\bf X}_5$ by taking $\omega = m / 2$ are the same form obtained in \eqref{cnserv-i-1}, \eqref{cnserv-i-2} and \eqref{cnserv-i-3}. The remaining conserved quantities associated with ${\bf X_6}$ and ${\bf X_7}$ given in \eqref{ns-class-I-s2} are
\begin{eqnarray}
    & & {\bf T}_6 =  F \, {\bf X}_6  + \frac{1}{F} \left[ \frac{H f_{,t}}{D}  \cos ( m t + \phi ) - f_{,r} \sin ( m t + \phi ) - \frac{f_{,\phi}}{D}  \cos ( m t + \phi ) \right] {\bf T}_0 , \quad \label{cnserv-cIsl-1} \\ & & {\bf T}_7 =  F \, {\bf X}_7 + \frac{1}{F} \left[ \frac{H f_{,t}}{D}  \sin ( m t + \phi ) + f_{,r} \cos ( m t + \phi ) - \frac{f_{,\phi}}{D}  \sin ( m t + \phi ) \right] {\bf T}_0 , \quad \label{cnserv-cIsl-2}
\end{eqnarray}
where $F$ and ${\bf T}_0$ are the same as Eqs. \eqref{defn-F} and \eqref{T0}, respectively. Further, the conservation law given in \eqref{con-law-3} for the Noether currents ${\bf T}_1,...,{\bf T}_5$ of the special class I case gives rise to the same equations with \eqref{cv-c13-1}, \eqref{cv-c13-2} and \eqref{cv-c13-3}. Additionally, for the Noether currents ${\bf T}_6$ and ${\bf T}_7$, the conservation law reads
\begin{eqnarray}
  & & F' + \frac{(H-1)}{D} F - f_{,t}^2 \frac{H}{F D} \left[ H + m (D^2 - H^2) \right] + f_{,r}^2 \left( \frac{D^2}{F} \right)'  \nonumber \\ & &  \qquad \qquad  \qquad \qquad \,\,\, + f_{,\phi}^2 \frac{(m H -1)}{F D} + \frac{ f_{,t} f_{,\phi}}{F D} \left[ 2 H + m (D^2 - 2 H^2) \right] = 0 \, , \label{cv-sc1assI-1} \\ & & f_{,r} f_{,t} \left[ \left( \frac{H D}{F} \right)' + \frac{H}{F} + \frac{m (D^2 - H^2)}{F} \right] - f_{,r} f_{,\phi} \left[ \left( \frac{D}{F} \right)' + \frac{1}{F} - \frac{m H}{F} \right] = 0 \, . \quad  \label{cv-sclassI-2}
\end{eqnarray}
For the special class I, three distinct cases arise from the conservation relations delineated in class I. While the outcomes derived in class I remain applicable to the special class I, Eq. \eqref{cv-sc1assI-1} further yields $m=1$ in case ({\it c}).

For the class II, where $H(r) = - \omega\, r^2$ and $D(r) = r$, the components of the symmetry generator ${\bf X}$ and the gauge vector ${\bf A}$ are obtained as
\begin{eqnarray}
& & \eta = c_1, \quad \xi^1 = \omega \, r (c_2 \sin \phi - c_3 \cos \phi) + c_4 , \,\, \nonumber \\& & \xi^2 =  c_2 \cos \phi + c_3 \sin \phi, \quad \xi^3 = \frac{1}{r} (-c_2 \sin\phi + c_3 \cos \phi ) + c_5 \, , \nonumber \\ & & A^1 = \lambda \, \omega\, r^2 f ( c_2 \sin\phi - c_3 \cos \phi ) + c_4 \, \lambda \, r f  + \Phi^1(t,r,\phi), \\ & &  A^2 = \lambda \, r f \, (c_2 \cos\phi + c_3 \sin\phi) + \Phi^2 (t,r,\phi), \nonumber \\ & & A^3 = c_1 \lambda \, r \,\phi + \lambda \, f ( -c_2 \sin \phi + c_3 \cos\phi + c_5 \, r )  - \int{(\Phi^1_{,t} + \Phi^2_{,r} ) d \phi } + \Phi^3 (t,r) \, , \nonumber
\end{eqnarray}
which yields that the {\it five KVs} ${\bf K}_1, {\bf K}_2, {\bf K}_3, {\bf K}_4$ and ${\bf K}_5$ given in \eqref{KV-class-II} are equivalent to the NSs ${\bf X}_1, {\bf X}_2, {\bf X}_3, {\bf X}_4$ and ${\bf X}_5$, respectively. Afterwards, we can write that the corresponding gauge vector fields are of the form
\begin{eqnarray}
  & & {\bf A}_1 = \lambda \, r \, f  \, {\bf X}_1  \, , \qquad {\bf A}_2 = \lambda \, r \, \phi \, {\bf X}_3 \, , \quad \,\, {\bf A}_3 = \lambda \, r \, f \, {\bf X}_3 \, , \label{gauge-classII-1} \\ & & {\bf A}_4 = \lambda \, r \, f \, {\bf X}_4 \, , \qquad   {\bf A}_5 =  \lambda \, r \, f \, {\bf X}_5 \, .  \label{gauge-i-2}
\end{eqnarray}
The Lie algebra of NSs for the class II will have the following non-vanishing commutators:
\begin{eqnarray}
& & \left[ {\bf X}_3, {\bf X}_4 \right] = {\bf X}_5, \quad \left[
{\bf X}_3, {\bf X}_5 \right] =  {\bf X}_4, \quad \left[ {\bf
X}_4, {\bf X}_5 \right] = 2 \omega {\bf X}_1 . \nonumber
\end{eqnarray}
Hence, the conserved vector fields associated with Noether symmetries are
\begin{eqnarray}
  & & {\bf T}_1 = F \, {\bf X}_1 + \frac{r^2}{F} f_{,t} {\bf T}_0 \, ,\,\, {\bf T}_2 = - \lambda \, r \, \phi {\bf X}_3 - \frac{r^2}{F} {\bf T}_0 \, ,\,\, {\bf T}_3 = F \, {\bf X}_3 + \frac{r^2}{F} f_{,\phi} {\bf T}_0 \, ,  \label{cv123-classII} \\ & & {\bf T}_4 = F \, {\bf X}_4 - \frac{r^2}{F} \left[ \omega \, r f_{,t} \sin\phi + f_{,r} \cos\phi - \frac{1}{r} f_{,\phi} \sin\phi \right] {\bf T}_0 \, , \label{cv4-classII} \\ & & {\bf T}_5 = F \, {\bf X}_5 - \frac{r^2}{F} \left[ \omega \, r f_{,t} \cos\phi - f_{,r} \sin\phi - \frac{1}{r} f_{,\phi} \cos\phi \right] {\bf T}_0 \, , \label{cv5-classII}
\end{eqnarray}
where $F$ and ${\bf T}_0$ are of the form
\begin{eqnarray}
  & & F = r \sqrt{ 1 + ( 1- \omega^2 r^2 )f^2_{,t} - 2 \omega f_{,t} f_{,\phi} - f_{,r}^2 - \frac{1}{r^2} f^2_{,\phi}} \, , \label{F-classII} \\ & & {\bf T}_0 = \left[ (\omega^2 r^2 -1) f_{,t} + \omega f_{,\phi} \right] \partial_t + f_{,r} \partial_r + \left( \omega f_{,t} + \frac{1}{r^2} f_{,\phi} \right) \partial_{\phi} \, . \label{T0-classII}
\end{eqnarray}
Then, applying the conservation law ${\bf D}.{\bf T} = 0$ to the conserved vector fields in \eqref{cv123-classII}, \eqref{cv4-classII} and \eqref{cv5-classII}, we found the following constraint equations:
\begin{eqnarray}
  & & f_{,t} \, f_{,r} \left( \frac{r^2}{F} \right)' = 0 \, , \quad  f_{,r} f_{,\phi} \left( \frac{r^2}{F} \right)' = 0 \, , \quad f_{,r} \left( \frac{r^2}{F} \right)' + \lambda \, r = 0 \, , \label{cv-classII-1} \\ & & F' - \frac{1}{r} F + \frac{\omega^2 r^3}{F} f_{,t}^2 + f_{,r}^2 \left( \frac{r^2}{F} \right)' - \frac{f_{,\phi}^2}{r\, F}  = 0 \, , \label{cv-classII-2}  \\ & &  \omega f_{,r} f_{,t} \left[ \left( \frac{r^3}{F} \right)' - \frac{r^2}{F} \right] - f_{,r} f_{,\phi} \left[ \left( \frac{r}{F} \right)' + \frac{1}{F} \right] = 0 \, . \label{cv-classII-3}
\end{eqnarray}
For class II, three distinct scenarios arise from the equations provided in \eqref{cases-abc} as follows:
\begin{eqnarray}
  & & (a) \, \left( \frac{r^2}{F} \right)' = 0 \, , \, \lambda = 0; \quad (b)\, f_{,r} = 0 \, , \, \lambda = 0;\quad (c)\, f_{,t} = f_{,\phi} = 0\, , \, \lambda \neq 0 \, . \label{cases-abc-II}
\end{eqnarray}
In case ({\it a}), we derive a solution $F= r^2/h (f_{_{,t}}\, , f_{_{,r}}\, , f_{_{,\phi}})$, where $h (f_{_{,t}}\, , f_{_{,r}}\, , f_{_{,\phi}})$ is an integration function. Consequently, Eq. \eqref{cv-classII-2} reads
\begin{eqnarray}
  & & f_{,\phi} = 0 \, , \qquad \omega^2 h^2 f_{,t}^2 + 1 = 0 \, , \label{ceq-1-classII}
\end{eqnarray}
while Eq.\eqref{cv-classII-3} becomes an identity. By substituting the relation $F = r^2 / h$ into \eqref{F-classII} and employing \eqref{ceq-1-classII}, after some algebraic manipulation, we ascertain that $f_{,r}^2 - f_{,t}^2 - 1 = 0$ and $F = \pm \, \omega \, r^2 \sqrt{1 - f_{,r}^2}$. In case ({\it b}), all constraint equations \eqref{cv-classII-1}-\eqref{cv-classII-3} become identities. In case ({\it c}), no solution satisfying the constraint equations is found.

\subsection{Class IV}

In this class, where $m^2 \neq 0, \omega = 0$, the metric functions are taken as $H(r) = 0$ and $D(r) = \frac{1}{m} \sinh(m r)$ for $m^2 >0$, or $D(r) = \frac{1}{\mu} \sin(\mu r)$ for $\mu^2 = -m^2 >0$.
The general solution to the Noether symmetry equations \eqref{ns-eqs} for this class gives
\begin{eqnarray}
& & \eta = c_1 - c_2 t, \quad \xi^1 = c_2 f + c_3  \, , \,\, \xi^2 =  c_4 \sin\phi + c_5 \cos\phi \, , \nonumber  \\ & &  \xi^3 = \frac{D'}{D} (c_4 \cos\phi - c_5 \sin\phi) + c_6  \, , \quad A^1 = \lambda \, f D ( c_2 f + c_3) + \Phi^1(t,r,\phi), \label{ns-class-4} \\ & &  A^2 = \lambda \, f D \, (c_4 \sin\phi + c_5 \cos\phi) + \Phi^2 (t,r,\phi), \nonumber \\ & & A^3 = \lambda \, \phi D ( c_1 - c_2 t) +  \lambda \, f  D' ( c_4 \cos\phi - c_5 \sin\phi ) + c_6 \lambda \, f D \nonumber \\ & & \qquad \quad - \int{(\Phi^1_{,t} + \Phi^2_{,r} ) d \phi } + \Phi^3 (t,r) \, . \nonumber
\end{eqnarray}
where $c_i$'s are constant parameters. So, one can write that there are {\it six} NSs given by
\begin{eqnarray}
  & & {\bf X}_1 = {\bf K}_1 \, , \quad {\bf X}_2 = {\bf K}_2 \, , \quad {\bf X}_3 = f \partial_t - t \partial_f = {\bf K}_3 - 2 t {\bf K}_2 \, , \\ & &  {\bf X}_4 = {\bf K}_4 \, , \quad {\bf X}_5 = {\bf K}_5 \, , \quad {\bf X}_6 = {\bf K}_6 \, ,
\end{eqnarray}
with the corresponding gauge vectors
\begin{eqnarray}
  & & {\bf A}_1 = \lambda \, f D \, {\bf X}_1 \, , \quad \qquad {\bf A}_2 = \lambda \, \phi D \, {\bf X}_6 \, , \quad \qquad {\bf A}_3 = \lambda D \, ( f^2, 0, - t \phi ) \, , \\ & & {\bf A}_4 = \lambda \, f D \, {\bf X}_4 \, , \quad \qquad {\bf A}_5 = \lambda \, f D \, {\bf X}_5 \, , \quad \qquad {\bf A}_6 = \lambda \, f D \, {\bf X}_6 \, ,
\end{eqnarray}
where the KVs ${\bf K}_1, ..., {\bf K}_6$ for this class are given in \eqref{KV-case-ii}.
Here, the vector fields ${\bf X}_1,{\bf X}_2,{\bf X}_4,{\bf X}_5$ and ${\bf X}_6$ given above have the same form with the corresponding KVs. The non-vanishing commutators of the NSs are
\begin{eqnarray}
& & \left[ {\bf X}_1, {\bf X}_3 \right] = -{\bf X}_2 \, , \qquad \left[
{\bf X}_2, {\bf X}_3 \right] = {\bf X}_1 \, , \nonumber \\& & \left[
{\bf X}_4, {\bf X}_5 \right] = - m^2 {\bf K}_6 , \quad \left[ {\bf X}_4, {\bf X}_6 \right] = -{\bf X}_5 \, , \quad \left[{\bf X}_5, {\bf X}_6 \right] = {\bf X}_4  \,  . \nonumber
\end{eqnarray}
Furthermore, the conserved vector fields for each NSs of this class read
\begin{eqnarray}
  & & {\bf T}_1 = F \, {\bf X}_1 - \frac{1}{F} f_{,t} {\bf T}_0 \, , \quad {\bf T}_2 = - \lambda \phi D \, {\bf X}_6 + \frac{1}{F} {\bf T}_0 \, , \label{cv12-classIV} \\ & &  {\bf T}_3 = f F {\bf X}_1 +  \lambda\ t\, \phi D \, {\bf X}_6 - \frac{1}{F} ( t + f f_{,t} ) {\bf T}_0 \, , \label{cv3-classIV} \\ & &  {\bf T}_4 = F \, {\bf X}_4 - \frac{1}{F} \left( f_{,r} \cos\phi - \frac{D'}{D} f_{,\phi} \sin\phi \right) {\bf T}_0 \, , \label{cv4-classIV} \\ & & {\bf T}_5 = F \, {\bf X}_5 - \frac{1}{F} \left( f_{,r} \sin\phi + \frac{D'}{D} f_{,\phi} \cos\phi \right) {\bf T}_0 \, , \,\, {\bf T}_6 = F \, {\bf X}_6 - \frac{1}{F} f_{,\phi} {\bf T}_0 \, , \label{cv5-classIV}
\end{eqnarray}
where ${\bf T}_0 = D^2 f_{,t} \partial_t - D^2 f_{,r} \partial_r - f_{,\phi} \partial_{\phi}$ and the function $F$ given by \eqref{defn-F} becomes
\begin{eqnarray}
  & & F = \sqrt{D^2 \left( 1 + f_{,t}^2 - f_{,r}^2 \right) - f_{,\phi}^2} \, . \label{F-class4}
\end{eqnarray}
One can eventually apply the conservation law \eqref{con-law-3} for the Noether currents obtained in \eqref{cv12-classIV}-\eqref{cv5-classIV}, and find the the following relations:
\begin{eqnarray}
    & & f_{,r} f_{,t} \left( \frac{D^2}{F} \right)' = 0 \, , \qquad f_{,r} \left( \frac{D^2}{F} \right)' + \lambda D = 0 \, ,  \label{cv-classIV-1} \\ & & F' - \frac{D'}{D} F +  f_{,r}^2 \left( \frac{D^2}{F} \right)' - f_{,\phi}^2 \frac{D'}{F D}  = 0 \, ,  \quad  f_{,r} f_{,\phi} \left[ \left( \frac{D D'}{F} \right)' + \frac{1}{F} \right] = 0 \, , \label{cv-classIV-2} \\ & & f_{,r} ( f f_{,t} + t) \left( \frac{D^2}{F} \right)' + f_{,t} \frac{2 D^2}{F}  +  \lambda \, t D = 0 \, .  \label{cv-classIV-3}
\end{eqnarray}

From the above equations, it is evident that this class also exhibits the same three possibilities outlined in \eqref{cases-abc}. In case (a), we offer solutions for the constraint equations \eqref{cv-classIV-1}-\eqref{cv-classIV-3} as follows:
\begin{eqnarray}
  & & F = \frac{D^2}{h} \, , \quad  f_{,t} = 0 \, , \quad f_{,r}^2 = \pm \sqrt{1 -  \frac{2 D^2}{h^2}} \, , \quad f_{,\phi} = \pm \frac{D^2}{h} \, ,
\end{eqnarray}
where $h = h (f_{_{,t}}\, , f_{_{,r}}\, , f_{_{,\phi}})$. In case ({\it b}), Eq.\eqref{cv-classIV-3} results $f_{,t} = 0$, indicating $f$ as a function of $\phi$ only. Consequently, the function $F$ defined in \eqref{F-class4} reduces to $F= \sqrt{D^2- f_{,\phi}^2}$. When considering this $F$ in \eqref{cv-classIV-2}, both equations are identically satisfied. For case ({\it c}), Eqs. \eqref{cv-classIV-1}-\eqref{cv-classIV-3} yield the same equations as \eqref{cv-c13-1-2} and \eqref{cv-c13-3-2}. Hence, the solutions for these equations remain the same as provided in \eqref{sol-fr-2} for $m^2 >0$ and $\mu^2 = - m^2 >0$, respectively.


\section{Conclusions}
\label{conc}

In this study, we employed the classical Noether symmetry approach to derive symmetries of minimal surface Lagrangian and the conserved vector fields corresponding to the Noether symmetry generators for G\"{o}del-type spacetimes. Using the minimal surface Lagrangian $L$ obtained from G\"{o}del-type spacetimes, we found the Noether symmetries for classes I, II, and III, resulting in {\it five} NS generators. Consequently, G\"{o}del-type spacetimes corresponding to these classes exhibit the algebra $\mathcal{N}_5$, equivalent to the Killing algebra $\mathcal{G}_5$. Notably, in special class I (where  $m^2 = 4 \omega^2$) and class IV, we identified {\it seven} and {\it six} NS generators, respectively. The special class I admits the algebra $\mathcal{N}_7 \equiv \mathcal{G}_7$, while class IV spacetime admits the algebra  $\mathcal{N}_6 \equiv \mathcal{G}_6$.
We observed that the Lagrangian for minimizing the 3-area enclosing a constant three-volume in G\"{o}del-type spacetimes shares a Lie algebra of Noether symmetries identical to the Lie algebra of KVs of those spacetimes. The corresponding gauge vectors are proportional to the KVs, which are equivalent to the Noether symmetries, although they do not mirror the exact form to the symmetries proposed in Ref.
\cite{qadir2012}.
Moreover, we determined the Noether currents resulting from the existence of Noether symmetries for minimal surface Lagrangians across each class I, II, III, and IV. Using the obtained Noether currents for all classes of G\"{o}del-type spacetimes, we applied the conservation law \eqref{con-law-3} to derive the analytical solutions of minimal surface equation \eqref{EL-eq-lagr}, showcasing the practicality and efficacy of Noether symmetries in this context.






\begin{thebibliography}{99}

\bibitem{godel} K. G\"{o}del, ``An example of a new type of cosmological solution of Einstein's field equations of gravitation,'' Rev. Mod. Phys. \textbf{21}(1949) 447.

\bibitem{kramer} H. Stephani, D. Kramer, M. A. H. MacCallum, C. Hoenselaers and E. Herlt,
``Exact Solutions of Einstein Field Equations,'' Cambridge University Press, (2003).

\bibitem{rayc} A. K. Raychaudhuri and S. N. Thakurta, ``Homogeneous space-times of G\"{o}del type,'' Phys. Rev. D \textbf{22} (1980) 802.
\bibitem{rebo1} M. J. Rebou\c{c}as and J. Tiomno, ``Homogeneity of Riemannian space-times of G\"{o}del type,'' Phys. Rev. D \textbf{28} (1983) 1251.
\bibitem{calvao1} M. O. Calv\~{a}o, M. J. Rebou\c{c}as, A.F.F Teixeira and W.M. Silva, ``Notes on a class of homogeneous space-times,'' J. Math. Phys. \textbf{\bf 29} (1988) 683.
\bibitem{teix} A. F. F. Teixeira, M. J. Rebou\c{c}as and J. E. \AA man, ``Isometries of homogeneous G\"{o}del type spacetimes,'' Phys. Rev. D \textbf{32} (1985) 3309.
\bibitem{rebo2} M. J. Rebou\c{c}as and J. Tiomno, ``A class of inhomogeneous G\"{o}del-type models,''  Nuova Cimento B \textbf{90} (1985) 204.
\bibitem{rebo3} M. J. Rebou\c{c}as and J. E. \AA man, ``Computer-aided study of a class of Riemannian space-times,'' J. Math. Phys. \textbf{28} (1987) 888.

\bibitem{qadir2012} A. Aslam and A. Qadir, ``Noether Symmetries of the Area-Minimizing
Lagrangian,'' {J. Appl. Math.} \textbf{2012} (2012)  532690.

\bibitem{tpq2015} M. Tsamparlis, A. Paliathanasis and A. Qadir,  ``Noether symmetries and
isometries of the minimal surface Lagrangian under constant volume in a Riemannian space,'' Int. J. Geom. Methods Mod. Phys. \textbf{12} (2015) 1550003.

\bibitem{mhs2023} A. Mohammadpouri, M. S. Hashemi  and S. Samaei, ``Noether Symmetries and Isometries of the Area-Minimizing Lagrangian on Vacuum Classes of pp-waves,'' {Eur. Phys. J. Plus} \textbf{138} (2023) 112.
\bibitem{js2016} S. Jamal and G. Shabbir, ``Noether symmetries of vacuum classes of pp-waves and the wave equation,'' {Int. J. Geometr. Meth. Modern Phys.} \textbf{13} (2016) 1650109.

\bibitem{camci2014} U. Camci, S. Jamal and A. H. Kara, ``Invariances and Conservation Laws Based on Some FRW Universes,'' {Int. J. Theor. Phys.} \textbf{53} (2014) 1483.

\bibitem{kg-godel} B. D. B. Figueiredo, I. D. Soares and J. Tiomno, ``Gravitational coupling of Klein-Gordon and Dirac particles to matter vorticity and spacetime torsion,'' {Class. Quantum Grav.} \textbf{8} (1992) 1593.

\bibitem{marecki} P. Marecki, ``On the wave equation in spacetimes of G\"{o}del type, chapter in G\"{o}del type spacetimes: history and new developments'', Annals of the Kurt G\"{o}del Society, ed. M. Scherfner and M. Plaue (2010); arXiv preprint gr-qc/0703018.

\bibitem{tp2010} M. Tsamparlis and A. Paliathanasis, ``Lie and Noether symmetries of geodesic equations and collineations,'' {Gen. Rel. Grav.} \textbf{42} (2010) 2957. 

\bibitem{tp2011} M. Tsamparlis and A. Paliathanasis, ``The geometric nature of Lie and Noether symmetries,'' {Gen. Rel. Grav.} \textbf{43} (2011) 1861.

\bibitem{cy2014} U. Camci and A. Yildirim, ``Lie and Noether symmetries in some classes of pp-wave spacetimes,'' {Phys. Scr.} \textbf{89} (2014) 084003

\bibitem{feroze1} T. Feroze, F.M. Mahomed and A. Qadir, ``The connection between isometries and symmetries of geodesic equations of underlying spaces,'' {Nonlinear Dynam.} \textbf{45} (2006) 65.

\bibitem{feroze2} T. Feroze, ``New conserved quantities for the spaces of different curvatures,'' {Modern Phys.Lett.} \textbf{A25} (2010) 1107.

\bibitem{feroze3} T. Feroze and I. Hussain, ``Noether symmetries and conserved quantities for spaces with a section of zero curvature,'' {J. Geom. Phys.} \textbf{61} (2011) 658.

\bibitem{feroze4} F. Ali and T. Feroze, ``Classification of plane symmetric static space-times according to their Noether symmetries,'' {Int. J. Theor. Phys.} \textbf{52} (2013) 3329.

\bibitem{cy2015}  U. Camci and A. Yildirim, ``Noether gauge symmetry classes for pp-wave spacetimes,'' {Int. J. Geom. Methods Mod. Phys.} \textbf{12}  (2015) 1550120.

\bibitem{ugur} U. Camci, ``Dirac analysis and integrability of geodesic equations for cylindrically symmetric spacetimes,'' {Int. J. Mod. Phys.} \textbf{12} (2003) 1431.

\bibitem{kundt} W. Kundt, ``Tr\"{a}gheitsbahnen in einem von G\"{o}del angegebenen Modell,'' {Z. Phys.} \textbf{145} (1956) 661.

\bibitem{chandra} S. Chandrasekhar and J.P. Wright, ``The geodesics in G\"{o}del's universe,'' {Proc. Natl. Acad. Sci.} \textbf{47} (1961) 341.

\bibitem{novello} M. Novello, I. D. Soares and J. Tiomno, ``Geodesic motion and confinement in G\"{o}del universe,'' {Phys. Rev. D} \textbf{27} (1983) 779.

\bibitem{rebo4} M. J. Rebou\c{c}as and A. F. F. Teixeira, ``Features of a relativistic space-time with seven isometries,'' {Phys. Rev. D} \textbf{34} (1986) 2985.

\bibitem{som} M. M. Som and A. K. Raychaudhuri, ``Cylindrically Symmetric Charged Dust Distributions in Rigid Rotation in General Relativity,'' {Proc. Roy. Soc. London} \textbf{A 304} (1968) 81.

\bibitem{paiva} F. M. Paiva, M. J. Rebou\c{c}as and A.F.F Teixeira, ``Time-travel in the homogeneous Som-Raychaudhuri universe,'' {Phys. Lett. A} \textbf{126} (1987) 168.

\bibitem{grave} F. Grave, M. Bauser, T. M\"{u}ller, G. Wunner and W.P. Schleich, ``The G\"{o}del universe:Exact geometrical optics and analytical investigation on motion,'' {Phys. Rev. D} \textbf{80} (2009) 103002.

\bibitem{kajari} E. Kjari, R. Walseri W. P. Schleich and A. Delgado, ``Sagnac effect of G\"{o}del's universe,'' {Gen. Rel. Grav.} \textbf{36} (2004) 2289.

\bibitem{daut} G. Dautcourt, ``The lightcone of G\"{o}del-like spacetimes,'' {Class. Quant. Grav.} \textbf{27} (2010) 225024.

\bibitem{calvao2} M. O. Calv\~{a}o, I.D. Soares and J. Tiomno, ``Geodesics in G\"{o}del-type space-times,'' {Gen. Rel. Grav.} \textbf{22} (1990) 683.

\bibitem{gleiser}  R.J. Gleiser, M. G\"{u}rses, A. Karasu and \"{O}. Sar{\i}o\~{g}lu, ``Closed timelike curves and geodesics of G\"{o}del-type metrics,'' {Class. Quantum Grav.} \textbf{23} (2006) 2653.

\bibitem{godel2014} U. Camci, ``Symmetries of geodesic motion in G\"{o}del-type spacetimes,'' {JCAP} \textbf{07} (2014) 002.

\bibitem{camci2015} U. Camci, ``Noether gauge symmetries of geodesic motion in stationary and
nonstatic G\"{o}del-type spacetimes,'' {Int. J. Mod. Phys.} \textbf{38} (2015) 1560072.

\bibitem{hall-costa} G. S. Hall and J. da Costa, ``Affine collineations in space-times,'' {J. Math. Phys.} \textbf{29} (1988) 2645.

\bibitem{melfo} A. Melfo, L. A. Nunez, U. Percoco and V. M. Villalba, ``Collineations of G\"{o}del-type space-times,'' {J. Math. Phys.} \textbf{33} (1992) 2258.

\bibitem{tsamparlis} M. Tsamparlis, D. Nikolopoulos and P. S. Apostolopoulos, ``Computation of the conformal algebra of 1+3 decomposable spacetimes,'' {Class. Quant. Grav.} \textbf{15} (1998) 2909.

\bibitem{cmc-sharif} U. Camci and M. Sharif, ``Matter collineations of spacetime homogeneous G\"{o}del-type metrics,'' {Class. Quant. Grav.} \textbf{20} (2003) 2169.

\bibitem{noether} E. Noether, ``Invariante Variationsprobleme,'' {G\"ottingen Math. Phys. Kl.} \textbf{2} (1918) 235; English translation by M.A. Tavel, ``Invariant Variation Problems,''
    {Transport Theory and Statistical Physics} \textbf{1}(3) (1971) 186.

\bibitem{anco1997} S. C. Anco and G. Bluman, ``Direct Construction of Conservation Laws from Field Equations,'' {Phys. Rev. Lett.} \textbf{78} (1997) 2869.

\end{thebibliography}
\end{document}